# PLASMA BRAIN DYNAMICS (PBD):
## A MECHANISM FOR EEG WAVES UNDER HUMAN CONSCIOUSNESS

John Z. G. Ma

ABSTRACT: EEG signals are records of nonlinear solitary waves in human brains. The waves have several types (e.g., $\alpha$, $\beta$, $\gamma$, $\theta$, $\delta$) in response to different levels of consciousness. They are classified into two groups: Group-1 consists of complex storm-like waves ($\alpha$, $\beta$, and $\gamma$); Group-2 is composed of simple quasilinear waves ($\theta$ and $\delta$). In order to elucidate the mechanism of EEG wave formation and propagation, this paper extends the Vlasov-Maxwell equations of Plasma Brain Dynamics (PBD) to a set of two-fluid, self-similar, nonlinear solitary wave equations. Numerical simulations are performed for different EEG signals. Main results include: (1) The excitation and propagation of the EEG wave packets are dependent of electric and magnetic fields, brain aqua-ions, electron and ion temperatures, masses, and their initial fluid speeds; (2) Group-1 complex waves contain three ingredients: the high-frequency ion-acoustic (IA) mode, the intermediate-frequency lower-hybrid (LH) mode, and, the low-frequency ion-cyclotron (IC) mode; (3) Group-2 simple waves fall within the IA band, featured by one or a combination of the three envelopes: sinusoidal, sawtooth, and spiky/bipolar. The study proposes an alternative model to Quantum Brain Dynamics (QBD) by suggesting that the formation and propagation of the nonlinear solitary EEG waves in the brain have the same mechanism as that of the waves in space plasmas.

KEYWORDS: Plasma brain dynamics (PBD); Quantum brain dynamics (QBD); Consciousness; EEG; nonlinear solitary wave

1. INTRODUCTION

Consciousness resides mainly in the outer layer of the cerebrum, cerebral cortex, with a thickness of $(2\sim5)\times10^{-3}$ m and a surface area of $0.16\sim0.4$ m$^2$,[1] giving a volume of $(3.2\sim20)\times10^{-4}$ m$^3$. The adult male human brain of an average of 1.5 kg has 86 billion neurons (nerve cells) and 85 billion non-neuronal cells.[2] The average volume density of neurons turns out to be $4.3\times10^{13} \sim 2.7\times10^{14}$ neurons /m$^3$.[3] Cortical neurons are interconnected with each other with each neuron to link with up to $10^4$ other neurons, forming a highly intricate system to pass signals via as many as 1000 trillion synaptic connections.[4]

Consciousness is dominated by the prefrontal cortex in the brain neuronal system to express the brain cognitive ability.[5] After the mathematical brain model was suggested in the early 1940s,[6] Quantum Brain Dynamics (QBD) was developed from Ricciardi and Urnezawa's pioneer work[7] to account for the neuro-and-cognitive mechanism of human consciousness.[8] QBD deals with a couple of basic fields: (1) the water ($H_2O$) rotational field, and (2) the electromagnetic field.[9] However, in spite of great progress having been achieved hitherto with sophisticated theories, models, or numerical simulations,[10] little advance has been reached toward understanding the measured low-frequency, nonlinear, electromagnetic brain waves in diagnosis of EEG (electroencephalography) or MEG (magnetoencephalography),[11] not mentioning the qualitative or quantitative data-fit visualizations of the holistic behavior of neuronal functions. To be more intriguingly, different from the pioneer recognitions in QBD, e.g., Penrose's neural "firing and not firing,"[12] and Penrose-Hameroff's neural microtubules,[13] new calculations of the neural de-coherence rates indicated that consciousness of the human brain should be thought of as a "classical rather than quantum" neural process, both for regular neuron firing and for kink-like polarization excitations in microtubules.[14] Therefore, it is necessary to review the physics in consciousness.

---

[1] Nunez PL, Srinivasan R. 2006. Electric fields of the brain: The neurophysics of EEG, 2nd ed. Oxford: Oxford University Press, p.6.

[2] Herculano-Houzel S. (1) 2009. The human brain in numbers: A linearly scaled-up primate brain. Front. Human Neurosci. 3, 31, pp.1-11; (2) 2016. The human advantage: A new understanding of how our brain became remarkable. Cambridge, MA: MIT Press, p.79.

[3] Teplan M 2002. Fundamentals of EEG measurement. Measurement Sci. Rev. 2, 2, pp.1-11.

[4] Mastin L 2010. Neurons & synapses. In: The human memory. http://www.human-memory.net/brain_neurons.html

[5] Gabi M, Neves K, Masseron C, et al 2016. No relative expansion of the number of prefrontal neurons in primate and human evolution. PNAS, 113, 34, 9617-9622.

[6] McCulloch WS, Pitts W 1943. A logical calculus of the ideas immanent in nervous activity. 5, pp.115-133; Reprint: 1990. Bull. Math. Biol. 52, 1/2, pp.99-115.

[7] Ricciardi LM, Umezawa H 1967. Brain and physics of many-body problems. Kybernetik, 4, 2, pp.44-48.

[8] Vitiello G 2011. Hiroomi Umezawa and quantum field theory. NeuroQuantology, 9, 3, pp.402-412.

[9] Jibu M, Yasue K. 1995. Quantum brain dynamics. In: Quantum brain dynamics and consciousness: An introduction. Amsterdam: John Benjamins Publishing, pp.163-166.

[10] E.g., (1) Başar E 2010. From quantum mechanics to the quantum brain. NeuroQuantology, 8, 3, pp.319-321. (2) Hameroff S 2012. How quantum brain biology can rescue conscious free will. Front Integr Neurosci. 6, 93, pp.1-17. (3) Sakane S, Hiramatsu T, Matsui T 2016. Neural network for quantum brain dynamics: 4D CP1+U(1) gauge theory on lattice and its phase structure. arXiv:1610.05443v1 [cond-mat.dis-nn].

[11] (1) For a detailed review, see: van Vugt M 2002. Oscillations in the brain: A dynamic memory model. Thesis, University College Utrecht. http://www.ai.rug.nl/~mkvanvugt/UCUthesis.pdf. (2)

Fortunately, an alternative model, Plasma Brain Dynamics (PBD), was proposed in the early 1970s by Hokkyo to deal with the collective features of the cortical phase transitions in neural populations of the living brain.[15] The preliminary study was based on plasma kinetics and made use of a quasi-linear approximation to the collisional Liouville equation to derive the distribution function, $F(\mathbf{r},\mathbf{v})$, of brain activities in the neural phase space (i.e., 3D position space $\mathbf{r}$ and 3D velocity space $\mathbf{v}$). Different from Ricciardi and Urnezawa's result that the stable memory of the brain contributed by the Bose condensation into the ground state of the quantum many-body system of Goldstone particles, Hokkyo concluded that a brain is (1) awaken but inattentive for a single-humped $F$ peaked at $\mathbf{v}=0$; (2) awaken but attentive for $F$ to possess a small second hump on the tail; and, (3) experiencing a memorizing process when the second hump is flattening to form a plateau which persists for indefinitely long time.

The brain plasma system can be treated as a collision-free media where neuronal activities could be mediated by long-range extracellular flows[16] and the collective behavior of the neuronal network could be described by equations of collision-free movements on the basis of stochastic analyses.[17] Such a system can be described by a set of Vlasov-Maxwell equations to evaluate brain functions by merely taking ion dynamics into consideration[18] in the nervous extracellular space where the interstitial fluid is in contact with the cerebrospinal fluid from the ventricular surfaces; surrounded by the extracellular space there exists larger intracellular neuronal compartment which occupies about 85% of the brain volume.[19] Because the extracellular electric signals, like microscopic-level spiking activity of neuronal assemblies, mesoscopic-level local field potential (LFP, also known as 'micro-EEG'), and macroscopic-level EEG, provide insights into the cooperative behavior of neurons, their average synaptic input and their spiking output,[20] we formulate collision-free plasma dynamics from Vlasov-Maxwell equations in this paper to focus on the mechanism of the EEG waves via data-fit modeling for the neuronal activity of the brain consciousness. It deserves a special mention here that any results of EEG analyses can be used for MEG due to the fact that the two types of signals have the same source of excitation, i.e., the ionic currents generated by biochemical processes at the cellular level, but recorded respectively in response to electric and magnetic effects of the brain activities.[21]

---

[17] Touboul J 2012. Mean-field equations for stochastic firing-rate neural fields with delays: Derivation and noise-induced transitions. Phys D: Nonlin Phenomena, 241, 15, pp.1223-1244.

[18] Tozzi A, Peters JF 2016. Towards plasma-like collisionless trajectories in the brain. viXra, 1610.0014v1, pp.1-10. http://vixra.org/pdf/1610.0014 v1.pdf

[19] Syková E, Nicholson C 2008. Diffusion in brain extracellular space. Physiol Rev, 88, 4, pp.1277-1340.

[20] Buzsáki G, Anastassiou CA, Koch C 2012. The origin of extracellular fields and currents -- EEG, ECoG, LFP and spikes. Nature Rev, Neurosci, 13, pp.407-420.

[21] da Silva FL 2013. EEG and MEG: Relevance to Neuroscience. Neuron, 80, 5, pp.1112-1128.

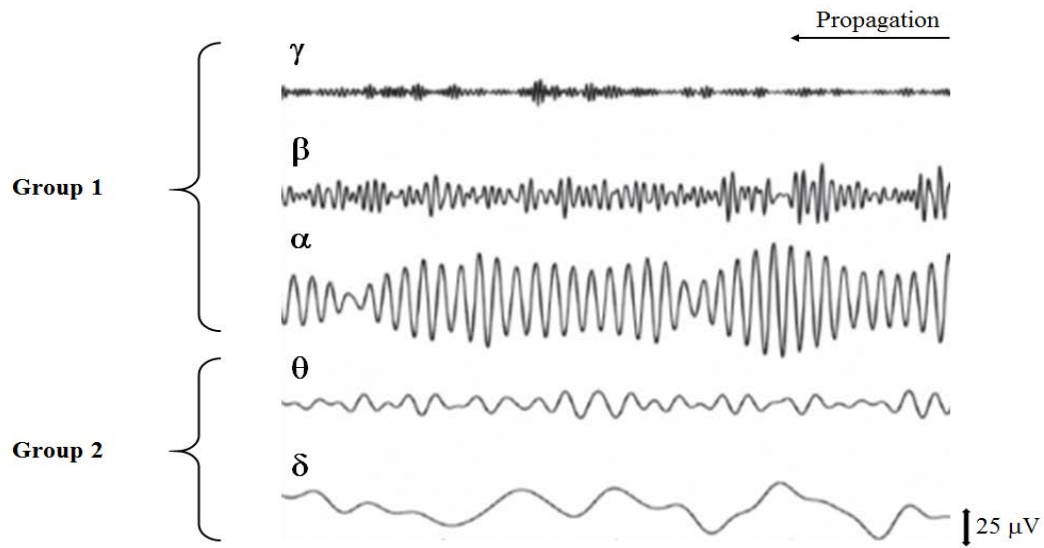

Figure 1. EEG signals in a rest state with closed eyes (adapted from Figure 2 in ref.22)

Figure 1 gives an example of the propagating EEG wave packages in a typical brain.[22] It exposes following features of consciousness:

(1) There exist five levels of consciousness (c.f., Thompson[23]): Semiconscious (bottom panel, $\delta$: 0.5-4 Hz; coma, dreamless-sleeping); Subconscious (lower middle panel, $\theta$: 4-8 Hz; drowsy, idling, dreaming, deep-meditation); Conscious (middle panel, $\alpha$: 8-14 Hz; relaxed, reflecting, light-meditation, visualization); Ultraconscious (upper middle panel, $\beta$: 14-30 Hz; perception, alerting, concentration); and, Superconscious (top panel, $\gamma$: 30-42 Hz; focus, religious ecstasy).

(2) Consciousness manifests itself extrinsically with nonlinear, solitary electromagnetic waves which are divided into a couple of groups: Group-1 consists of complex amplitude-modulated stormy waves[24] (as given in the upper three panels: $\alpha$, $\beta$, and $\gamma$) which are characterized by the superimposition of multi-frequency waveforms; and, Group-2 is of simple quasilinear waves[25] (as given in the lower two panels: θ and δ) which are characterized by envelopes between sinusoidal and saw-tooth waveforms of one-or-two dominant frequencies.

---

[22] Campisi P, La Rocca D, Scarano G 2012. EEG for automatic person recognition. Computer, 45, pp.87-89.

[23] Thompson E 2015. Dreamless Sleep, the embodied mind, and consciousness. In: T Metzinger & JM Windt (Eds). Open MIND, 37(T), Frankfurt am Main: MIND Group. doi: 10.15502/9783958570351, pp.1-19.

[24] Ma JZG, Hirose A 2010. Lower-hybrid (LH) oscillitons evolved from ion-acoustic (IA)/ion-cyclotron (IC) solitary waves: Effect of electron inertia. Nonlin. Proc. Geophys. 17, pp.245-268.

(3) The fluctuation of waveforms is approximately symmetric to the resting potential (0 μV), indicating that the brain functioning is bounded to, or, at least can be reduced to, a kind of two-species system of oppositely charged particles which are maintained by billions of neurons.[26] The dynamical interaction or competition of the two ingredients in the system pumps the excitation, development, and propagation of electric (or magnetic) signals which are able to be measured as EEG (or MEG) waveforms.

As the first study on the EEG mechanism through formulating the new PBD paradigm, the purpose of this paper lies in gaining substantial insights into the complicated features of the real-time EEG oscillations in realistic situations. The work will provide a reference to illustrate the synaptic, coherent propagating flows in the nervous extracellular space of the brain. No similar studies have yet been reported so far in literature. To reduce the complexity of the problem while still being able to develop a tenable approach toward the distinctive nonlinear characteristics of EEG signals of the brain, we assume a basic two-component plasma system consisting of positively and negatively charged ions. That is to say, all the ions with the same polarity in the brain are reduced to one-type, singly-charged species with a reduced mass, $m_i$, which is around tens of proton mass, $m_p$. In addition, we suggest that all of the test particles of the brain plasma under modeling are well inside the extracellular space thereby being able to neglect all the edge effects in a 3D Cartesian frame of reference. Furthermore, we neglect the relativistic effects of all the test particles considering the fact that the globally coherent speeds of the brain activity occurred in the five levels of consciousness are no more than $V_c \sim 5$ m/s.[27] specifically, the travelling speeds of the EEG signals were estimated with values, on average, of α: 6.5±0.9 m/s; θ: 4.0±0.9 m/s; as well as all of the signals fall within 3~11 m/s.[28]

Previous studies show that not only the axonal actions of the neuronal system are similar to the scaled equivalents of plasma lightning,[29] but also the cerebral cortex and its white matter system of corticocortical fibers turns out to be a system somewhat analogous to the earth's ionospheric shell.[30] We are thus inspired to extend our work on space plasma dynamics applicable to brain plasma dynamics, with the first step to elucidate the mechanism of EEG signals. The layout of the paper is as follows: Section 2 estimates brain plasma parameters and set up a set of two-fluid Vlasov-Maxwell equations; Section 3 derives a set of nonlinear, self-similar differential equations for the excitation and propagation of EEG waves; Section 4 exposes the data-fit modeling of the solitary EEG waves under different conditions. The last section gives conclusions along with some concise discussions. SI units are used throughout the paper except wherever the conventional units are more conveniently used.

---

[25] Ma JZG, Hirose A 2009. Parallel propagation of ion solitons in magnetic flux tubes. Phys. Scr., 79: 045502.

[26] Herculano-Houzel S 2009. The human brain in numbers: a linearly scaled-up primate brain. Front. Hum. Neurosci., 3, 31, pp.1-11.

[27] Alexander DM, Nikolaev AR, Jurica P, et al 2016. Global Neuromagnetic Cortical Fields Have Non-Zero Velocity. PLoS ONE, 11, 3, e0148413.

[28] Patten TM, Rennie CJ, Robinson PA, Gong P 2012. Human cortical traveling waves: Dynamical properties and correlations with responses. PLoS
ONE. 7, 6, e38392, pp.1-10.

[29] Persinger MA 2012. Brain electromagnetic activity and lightning: potentially congruent scale-invariant quantitative properties. Front Integr Neurosci, 6, 19, pp.1-7.

## 2. BRAIN PLASMA MODEL AND TWO-SPECIES VLASOV-MAXWELL EQUATIONS

Among the neurons of the cerebral cortex, there always exist transmissions of impulses which induce dendritic synapses to drive excitatory and inhibitory postsynaptic potentials. The triggered currents move through dendrites and cell body to the axon base, and pass through the membrane to the extracellular space. It was suggested that the superimposition of the potentials derived from the mixture of the extracellular currents generated by those neurons with uniformly oriented dendrites gives rise to EEG signals,[31] which has a characteristic magnitude of the electric field, $E_c$ = 2 mV/m for an order of ~200 μV with a typical distance of 10 cm.[32] The involved ions consist mostly of small ions like $H^+$, $Na^+$, $K^+$, $Ca^{2+}$, and $Cl^-$.[33]

In both the intracellular and extracellular spaces, the concentration of negative ions (124.0 mM) is far less than that of positive ones (317.5 mM), giving the charge number densities of $n_+ \approx 1.9 \times 10^{26}$ m$^{-3}$, and $n_- \approx 39\%$ $n_+$,[34] with $n_+ \sim$ 1/1000 of the molecular number density of water or the free electron density in copper. Because the brain is electrically neutral (i.e., as many positive charges as negative charges), the excess positive charges in the brain should be balanced by the abundant electrons which come from macromolecules such as nucleic acids and proteins in the brain (c.f. p.685 in ref.33). in this case, this paper assumes a model to the first order which treats the negative ions as a kind of "dust" ingredient in the brain, and neglect their contributions to reduce the complexity of the study, while still being able to obtain significant solutions toward elucidating the dominant EEG features as shown in Figure 1. With this simplification, we treat the brain plasma consists of negatively charged electrons only and singly charged one-species ion with a reduced mass $m_i$ from all of the ions.

On the one hand, the **E**×**B** drift of brain plasma particles in the typical external geomagnetic field, $B_{ext} \sim 5 \times 10^{-5}$ T, has a characteristic speed, $V_d = E_c / B_{ext}$ $\sim 40$ m/s, which applies for all particles including electrons, positive and negative ions. On the other hand, the brain aqua ions, $[M(H_2O)_n]^{z+}$ (where parameters M, n, and z are a metal atom, solvation number, and charge number, respectively), has a typical thermal speed, $V_{Ti} = \eta\sqrt{8k_b T_{i0}/\pi m_i}$ =105 m/s (where $k_b$ =1.38 $\times 10^{-23}$ J/K, $T_{i0}$ = 310 K, $m_i$ =148$m_p$=2.48×10$^{-25}$ kg ($m_p$ is the proton mass) for M=Ca and n=6; with $\eta \sim 0.5$, a relaxation coefficient due to de-coherence), while brain electrons are of $V_{Te} = \sqrt{8k_b T_{e0}/\pi m_e}$ =109 km/s for the same temperature $T_{i0}$. Thus, $V_d \sim V_{Ti} \ll V_{Te}$, suggesting that the electric field is more heavily correlated with the dynamics of ions rather than that of electrons. If $B_{ext}$ decreases and $V_d$ is competitive to $V_{Te}$, electron dynamics will certainly be different. This paper does not deal with large **E**×**B** drift case which is used for astronauts or those under clean magnetic environment.

The two types of particles are therefore described by different collision-free Vlasov-Maxwell equations. Reduced from the collisional Boltzmann equation, the mandatory set of the Vlasov-Maxwell equations for ions and electrons are expressed jointly as follows:[15,35]

$$\begin{cases} \left(\frac{\partial}{\partial t} + \mathbf{v} \cdot \nabla + \mathbf{a}_\alpha \cdot \nabla_\mathbf{v}\right) f_\alpha = 0 \\ \nabla \times \mathbf{E} = -\frac{\partial \mathbf{B}}{\partial t} \\ \nabla \times \mathbf{B} = \mu \mathbf{j} \\ \nabla \cdot \mathbf{E} = \frac{q}{\varepsilon} \\ \nabla \cdot \mathbf{B} = 0 \end{cases} \quad (1)$$

in which $t$ is time, subscript $\alpha = (i,e)$ denote ion and electron species, respectively; and

$$\mathbf{a}_\alpha = \frac{e_\alpha}{m_\alpha}(\mathbf{E} + \mathbf{v}\times\mathbf{B}) \ , \ q = \sum_\alpha e_\alpha \int f_\alpha \mathrm{d}\mathbf{v} \ , \ \mathbf{j} = \sum_\alpha e_\alpha \int \mathbf{v} f_\alpha \mathrm{d}\mathbf{v}$$

where $e_e = -e$ and $e_i = +e$. In addition, $\varepsilon = \varepsilon_r \varepsilon_0 \sim 80\varepsilon_0$ (p.118: ref.31) and $\mu = \mu_r \mu_0 \sim 10\mu_0$,[36] in which $\varepsilon_0$ and $\mu_0$ are the electric permittivity and magnetic permeability of free space, respectively.

---

[35] Boyd TJM, Sanderson JJ 2003. The physics of plasmas. Cambridge: Cambridge University Press. pp.253-255.

## 3. TWO-FLUID, SELF-SIMILAR NONLINEAR SOLITARY WAVE EQUATIONS IN BRAIN PLASMA

EEG waves are the collective manifestation of the brain plasma particles in the presence of both the internal brain electric & magnetic fields and the external geomagnetic field. The elucidation of the collective features depends on the fluid formulations of the brain plasma system. By introducing density $n$, velocity $\mathbf{u}$, and pressure $p$ (the diagonal value of tensor $\vec{\mathbf{P}}$) as follows:[37]

$$n_\alpha = \int f_\alpha \mathrm{d}\mathbf{v} \ , \ \mathbf{u}_\alpha = \frac{1}{n_\alpha} \int \mathbf{v} f_\alpha \mathrm{d}\mathbf{v} \ , \ \vec{\mathbf{P}}_\alpha = m_\alpha \int (\mathbf{v}-\mathbf{u})(\mathbf{v}-\mathbf{u}) f_\alpha \mathrm{d}\mathbf{v}$$

Equation (1) provides the following self-similar set of two-fluid equations for the brain plasma system by assuming a slab model where the propagation of EEG signals is along the $x$-coordinate in the Cartesian frame ($x,y,z$):

For the electron fluid:

$$\begin{cases} \left(u_{ex} - \dfrac{\xi_m}{u_{ex}}\right)\dfrac{du_{ex}}{dX} = -(u_{ey}B_z - u_{ez}B_y) \\ u_{ex}\dfrac{du_{ey}}{dX} = -(S_y - u_{ex}B_z + u_{ez}B_{x0}) \\ u_{ex}\dfrac{du_{ez}}{dX} = -(S_z + u_{ex}B_y - u_{ey}B_{x0}) \end{cases} \quad (2)$$

For the ion fluid:

---

[36] Georgiev DD 2003. Electric and magnetic fields inside neurons and their impact upon the cytoskeletal microtubules. Cogprints Report, U. Southampton, UK, Tech. Rep. http://cogprints.org/3190/

[37] Schunk RW, Nagy FA 2000. Ionospheres: physics, plasma physics, and chemistry. Cambridge: Cambridge University Press. pp.50-51.

$$\begin{cases} \xi_m \left(u_{ix} - \dfrac{3S_n^2}{\xi_T u_{ix}^3}\right)\dfrac{du_{ix}}{dX} = E_x + u_{iy}B_z - u_{iz}B_y \\ \xi_m u_{ix}\dfrac{du_{iy}}{dX} = S_y - u_{ix}B_z + u_{iz}B_{x0} \\ \xi_m u_{ix}\dfrac{du_{iz}}{dX} = S_z + u_{ix}B_y - u_{iy}B_{x0} \end{cases} \quad (3)$$

For electric and magnetic fields:

$$\begin{cases} \dfrac{dE_x}{dX} = -\xi_m n_{\text{sc}} \\ \left(1 - \dfrac{M^2\xi_V^2}{\xi_m}\right)\dfrac{dB_y}{dX} = S_n \xi_V^2 \left(\dfrac{u_{iz}}{u_{ix}} - \dfrac{u_{ez}}{u_{ex}}\right) \\ \left(1 - \dfrac{M^2\xi_V^2}{\xi_m}\right)\dfrac{dB_z}{dX} = -S_n \xi_V^2 \left(\dfrac{u_{iy}}{u_{ix}} - \dfrac{u_{ey}}{u_{ex}}\right) \end{cases} \quad (4)$$

in which

$$\begin{cases} S_n = u_{x0} - M, \ S_y = E_{y0} - MB_{z0}, \ S_z = E_{z0} + MB_{y0} \\ n_{\text{sc}} = n_e - n_i, \ n_e = \dfrac{S_n}{u_{ex}}, \ n_i = \dfrac{S_n}{u_{ix}} \\ E_y = S_y + MB_z, \quad E_z = S_z - MB_y, \quad B_x = B_{x0} \end{cases} \quad (5)$$

In Equations (2)~(5), subscript "$_0$" refers to the equilibrium state, and $X$ is the self-similar coordinate after the transformation of $X=x-Mt$,[38] where $M = V_{\text{ph}}/c_s = \sqrt{1 + \gamma/\xi_T^2} > 1$ is the Mach number which is independent of $X$ ($\gamma$ is the adiabatic index) and determined by $V_{\text{ph}}$ and $c_s = V_{Ti}\sqrt{\gamma\pi/8}/\eta$, the phase speed and the local sound speed of the ion acoustic (IA) waves [Eq.(10) in ref.39]. All of the parameters are dimensionless with normalizations of $n$ by $n_0$; $(x,y,z)$ by electron Debye length $\lambda_{\text{De}}$; $u$ by $c_s$; $p$ by $p_0$; $B$ by a pseudo-magnetic field $B_0=m_e\omega_{\text{pi}}/e$ ($\omega_{\text{pi}}$ is the ion plasma frequency, ); $E$ by $E_0=c_sB_0$. In addition, $\xi_m = m_i/m_e$ is ratio between ion and electron masses, $\xi_T = T_{e0}/T_{i0} = (p_{e0}n_{i0})/(p_{i0}n_{e0})$ is the ratio between electron and ion temperatures, and $\xi_V = v_{Te}\sqrt{\varepsilon_r\mu_r}/c \sim 28v_{Te}/c$ is the ratio between electron thermal speed and the speed of light in the brain, where $c$ is the speed of light in free space. Note that $\varepsilon_0 E_0^2/p_{e0} = (m_e/m_i)^2$.

This set of nonlinear equations was derived previously to describe the satellite-measured coherent solitary waves excited in the two-fluid system in space plasmas.[39] The only difference of the two-fluid formulations between the space plasma and brain plasma situations arise from the electric permittivity and magnetic permeability: in the former, the two parameters are those of the free space, $\varepsilon_0$ and $\mu_0$; by contrast, in the latter, they are updated by $\varepsilon$ and $\mu$. This fact brings us to make use of the data-fit modeling results in the classical ionospheric and magnetospheric physics to reexamine the physics of consciousness via EEG signals. The approach will thus provide an alternative paradigm to QBD in the study of the neural processes.

---

[38] Lee LC, Kan JR 1981. Nonlinear ion-acoustic waves and solitons in a magnetized plasma, Phys. Fluids 24, pp.430-436.

## 4. MECHANISM OF EEG WAVES

Under different electron conditions, Equations (2)~(5) describe both stormy EEG waves (Group 1: α, β, and γ) and simple EEG waves (Group 2: θ and δ) in Figure 1. The boundary conditions of the simulations at *X*=0 satisfy: $n_e \approx n_i = n_0$, $u_{ex} = u_{ix} = U_0 - M$, $E_y = E_{y0}$, $E_z = E_{z0}$, $B_x = B_{x0}$, $B_y = B_{y0}$, and $B_z = B_{z0}$. Note that in linear IA regime, once $\xi_T$ is given, *M* can be easily obtained; by contrast, in the nonlinear regime, the decoupled two parameters contribute jointly to the formation of solitary waves.

*4.1 Simple EEG waves (Group 2)*

Because the mass of the negatively charged electrons, $m_e = 9.1094 \times 10^{-31}$ kg, is far smaller than that of the positively charged ions, $m_i$, which is tens of the mass of protons, $m_p = 1836 m_e = 1.6726 \times 10^{-27}$ kg, it is reasonable to neglect the electron inertia in the complicated set of the nonlinear Equations (2)~(5) by assuming a zero electron mass. In this case, the dynamics of electrons are neglected and the simplest nonlinear solitary waves are obtained in the IA mode propagating in the direction parallel to the field lines of the external magnetic field, $B_{\text{ext}}$. A detailed analysis of the IA solitary waves was given in a little more complicated case,[40] where two types of isothermal Boltzmann electrons are included, one type is background electrons and the other is energetic ones. For the simpler case with only one type of background electrons, the reduced set of Equations (2)~(5) is as follows [Equation (9) in ref.39]:

---

[39] Ma JZG, Hirose A 2010. Lower-hybrid (LH) oscillitons evolved from ion-acoustic (IA) / ion-cyclotron (IC) solitary waves: effect of electron inertia. Nonlin. Processes Geophys. 17, pp.245–268.

[40] Ma JZG, Hirose A 2009. Parallel propagation of ion solitons in magnetic flux tubes, Phys. Scripta. 79, 045502, pp. 1-13.

$$\frac{d^2\varphi}{dX^2} = e^\varphi - \sqrt{\frac{2}{\left[1 + \frac{1}{M_*^2}\left(\frac{3}{\xi_T} - 2\varphi\right)\right] + \sqrt{\left[1 + \frac{1}{M_*^2}\left(\frac{3}{\xi_T} - 2\varphi\right)\right]^2 - \frac{12}{\xi_T M_*^2}}}} \quad (6)$$

in which $\varphi(X)$ is the normalized electrostatic potential in unit of $k_B T_{e0}/e$ to simulate the output of the EEG measurements, and $M_* = |U_0 - M|$. Notice that the output of EEG signals is the potential drop, $\phi(E) = \sum_n (d\varphi/dX)|_n \Delta X_n \approx (\sum_n E_n)d = Ed$, rather than the potential $\varphi(X)$, where $d$ is the full distance between the two electrodes to measure the drop.

Figure 2 presents an example of the $\phi(E)$ waveforms simulated under different sets of input parameters in the data-fit modeling. First of all, there exist three types of nonlinear solitary waves in view of the envelopes: sinusoidal (lower left panel), sawtooth (upper right panel), and spiky/bipolar (lower right panel). These nonlinear waveforms are developed from the non-wave structure (upper left panel) which is superimposed upon the background propagating linear IA waves and non-propagating IC oscillations (see Appendix B in ref.39 for the dispersion relations). Though the envelopes appear distinct among the three waves, previous FFT analyses have exposed that the dominant frequencies of the three solitary waves are all in the IA band, however, the number of harmonic ingredients is different from one type to the other. Take a look at Group 2 in Figure 1 and compare it with the three solitary waves in Figure 2. Not surprisingly, either θ or δ signals of the measured EEG waves turns out to be either one of the three types, or the composition of two or three of them. No doubt, it is evident that the simple EEG waves in Group 2 are the nonlinear IA solitary waves propagating in the brain.

Besides, there exist only two specific components of ions which are responsible for the excitation and propagation of the brain IA solitary waves. One component is at a lower initial subsonic speed, $U_{0l}$, which moves opposite to $X$, and the other one is at a higher initial supersonic speed, $U_{0h}$, which moves along $X$. In either cases, the difference between the two speeds and the transonic Mach number, $M$=1.04 (for γ=3 and $\xi_T = 5.8$), must be over-transonic with a value of $M_*$ to drive solitary waves. For $M_* = 1.3$, $U_{0l} = -0.26$ and $U_{0h} = 2.34$. Interestingly, as displayed in the two left panels of Figure 2, a small increase in $M_*$ from 1.3 to 1.4 propels the formation of the sinusoidal waves. In order words, there exists a minimum $|U_0|$ for either the subsonic or supersonic ions lower than which no solitary waves are able to be excited to produce EEG signals.

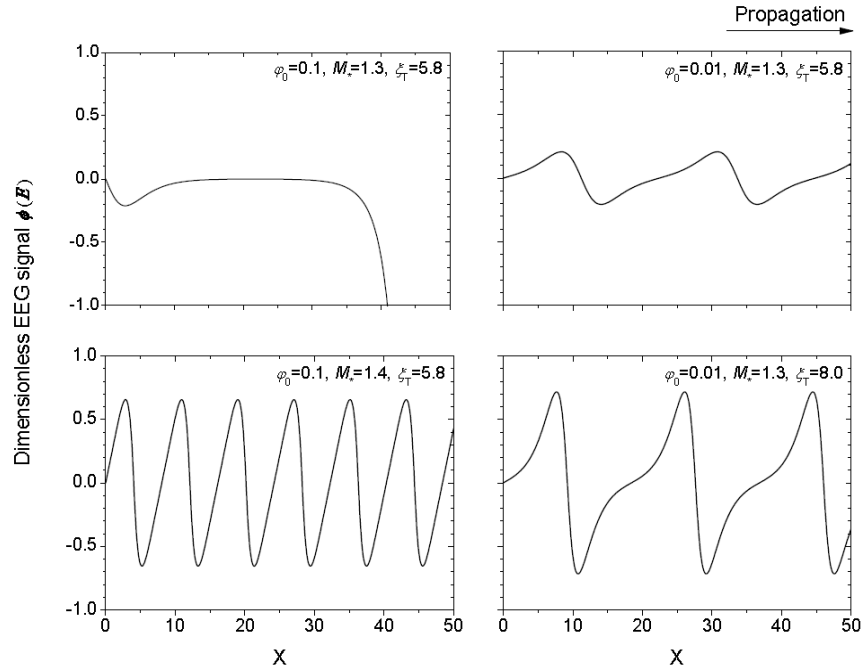

Figure 2. Simulations of simple EEG signals, $\phi(E)$

Furthermore, similar to the existence of the minimum $|U_0|$, there exist a maximum initial value in potential $\varphi(X)$, as shown in the upper two panels of the figure. At $\varphi_0=0.1$ no solitary waves appear for $M_* = 1.3$ and $\xi_T = 5.8$; when $\varphi_0$ decreases, coherent waveforms come into being. At $\varphi_0=0.01$ a train of sawtooth envelopes is developed to exhibit the solitary structures of propagating waves. In comparison, the ratio between electron and ion temperatures, $\xi_T$, also plays a role in the modulation of solitary waves. For example, a minimum ratio exists only above which can waves be driven; in addition, as given in the two right panels, the waves become spiky with the increase of the ratio. It deserves mentioning that the direction of the wave propagation in simulations is opposite to that of the measured signals due to the different frames of reference. However, there do remain oppositely propagating solitary waves simultaneously in the curvilinear coordinates owing to the presence of centrifugal and Coriolis factors.[41]

---

[41] Ma JZG 2010. Nonlinear ion-acoustic (IA) waves driven in a cylindrically symmetric flow. Astrophys Space Sci. 330, pp.87–94.

*4.2 Complex EEG waves (Group 1)*

In the simple EEG waves, electrons are assumed to response to the external electric and magnetic fields instantly to keep the plasma neutrality at any time. However, this assumption so idealizes the real situations as not to offer a validated mechanism to account for the complex EEG signals as illustrated in Group 1 of Figure 1. We have to relax the constraint on the electron inertia and take into account the role played by the electron mass. In this case, the dynamics of both ion and electron particles are involved in the nonlinear set of the highly coupled Eq. (2)~(5). No analytical solutions can be derived as that given in Eq. (6) for the simple EEG waves. Numerical simulations are required to reveal the solitary structures which should certainly be modulated by the newly embedded ingredient to display unknown solitary wave packets. These packets should be different from the three simple modes (sinusoidal, sawtooth, and spiky/bipolar) in the IA regime.

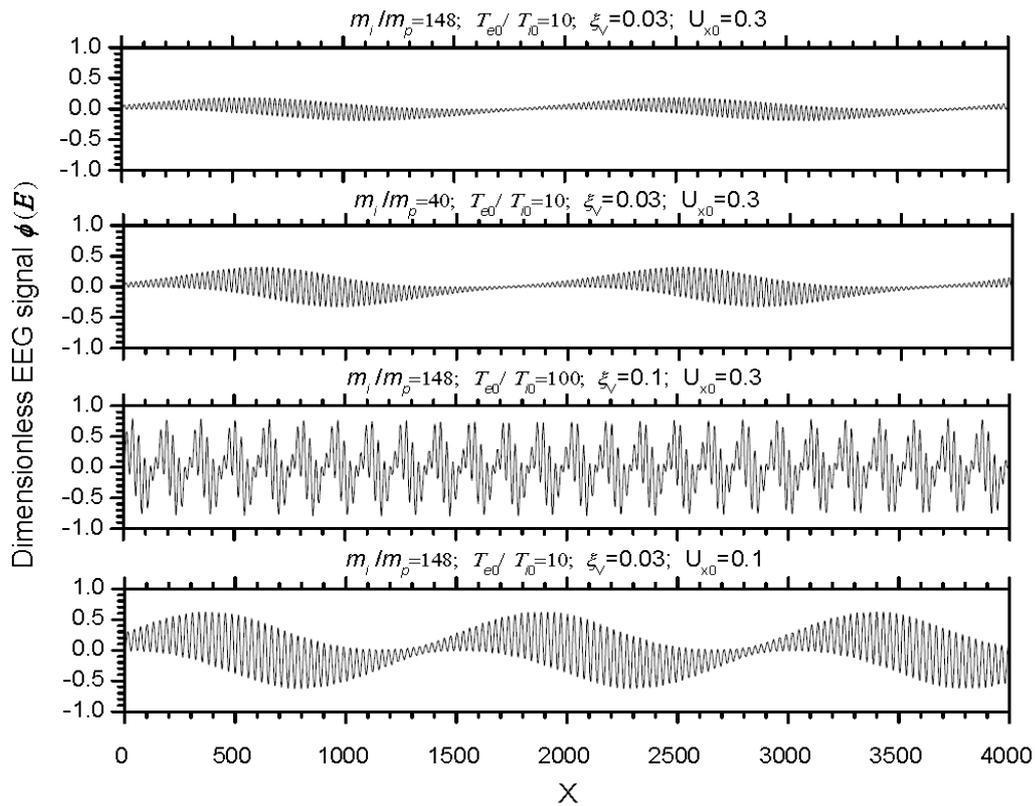

Figure 3. Simulations of complex EEG α-signals, $\phi(E)$

A typical set of input parameters used for simulations is taken as follows: $m_i/m_p = 148$, giving $\xi_m = 271728$; $\xi_T = T_{e0}/T_{i0} = 10$, giving $\xi_V = 0.03$; and $U_0$=(0.3,0,0), $M$=1, $E_0$=(0,0,1), $B_0$=(0.2,0,1). The results are given in Figures 3, 4, and 5 to expose the mechanism of how the EEG signals in Group-1, α, β, and γ, come into existence.

Firstly, Figure 3 illustrates the general modulations of the input parameters on the EEG envelopes with an emphasis on the EEG α-signals. These parameters under consideration in the present study include the ion & electron mass ratio, $m_i/m_p$, their temperature ratio, $T_{e0}/T_{i0}$, electron relative speed $\xi_V$, as well as the initial speeds of plasma fluids along $x$, $U_{x0}$. The top panel presents the wave packets under the condition restricted by the introduced typical set of input parameters. The FFT power spectra in space plasmas[39] shows that the oscillating frequencies contain following three bands: spiky high-frequency IA band, oscillating intermediate-frequency lower-hybrid (LH) band, and fluctuating low-frequency ion-cyclotron (IC) band. Though space and brain plasmas own different electric permittivity and magnetic permeability, the similar waveforms of the nonlinear solitary waves propagating in the two regions remind us at least qualitatively to assume that EEG signals have the same mechanism in these three wave bands. Quantitative studies are needed for the data-fit modeling in clinic applications. This is beyond the scope of the present study.

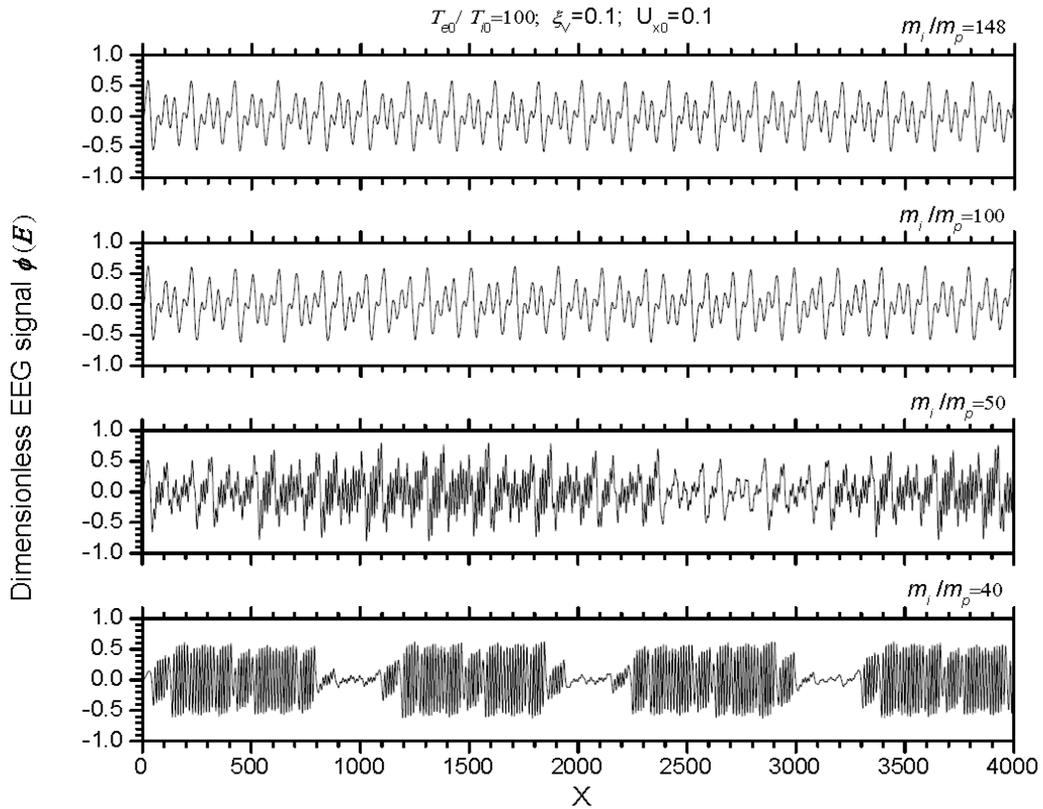

Figure 4. Simulations of complex EEG β-signals, $\phi(E)$

The top panel includes roughly two wave packets and 180 oscillations. The ratio of ~90 falls just in the range of that in IC and IA frequencies in ionospheric auroral F-layer.[42] With a reduced $m_i/m_p$ =40 in the upper middle panel, there is a little unnoticeable 6-period decrease in the oscillations, however, with an obvious increase in the magnitude of the wave packets from below 0.2 in the top panel to above 0.3 in this panel. When $T_{e0}/T_{i0}$ enhances from 10 to 100, however, the lower middle panel exposes a totally different train which contains 26 solitary packets and every packet owns 6 IA oscillations. These packets should be related to LH frequencies to which electrons are involved in the modulations, while the IC oscillations are undiscernible due to the long wave lengths. Notice that $\xi_V$ is dependent of $T_{e0}$. By contrast, a decrease in $U_{x0}$ from 0.3 in the top panel to 0.1 in the bottom panel reduces the IC wavelength and an additional packet appears.

---

[42] Burchill JK, Knudsen DJ, Bock BJJ, et al 2004. Core ion interactions with BBELF, lower hybrid, and Alfven waves in the high-latitude topside ionosphere, J. Geophys. Res. 109, A01219.

Secondly, since a brain contains ions with respective masses, Figure 4 particularly displays the impact of the mass on the solitary wave packets of $T_{e0}/T_{i0} = 100$ and $U_{x0}=0.1$, with an emphasis on the EEG β-signals. when $m_i/m_p$ decreases from 148 in the top panel to 100 in the upper middle panel, the IA-LH train of the 21 wave packages 21 in the top panel does not have obvious changes. By contrast, after the ratio reduces to 50 in the lower middle panel, the low-frequency IC feature emerges, along with approximately doubled high-frequency IA oscillations. Strikingly, with the ratio goes down to 40, meaning the dominance of the atomic $K^+$ and/or $Ca^{2+}$, over the aqua-ions in the brain plasma dynamics, a kind of new train of the oscillating solitary packets comes to birth, named as nonlinear "oscillitons" in space physics.[43] Oscillitons are embodied with a full spectra of oscillations from the lower IC end to the higher IA end. The frequency ratio in the case of the bottom panel is IC:LH:IA~1:10:60.

Finally, a direct comparison between Figure 1 and Figures 3 & 4 let us aware of the fact that the α-type EEG signal has similar appearances presented jointly by all of the panels in Figure 3; and the β-type signal does with all of the panels in Figure 4. Therefore, it is feasible to have a parameterized study to quantitatively determine realistic dimensional parameters through data-fit modeling. For the highly stormy γ-type EEG signal, the production of the highly nonlinear oscillitons in the bottom panel of Figure 4 guides us to believe that only low mass ions are responsible for the origin of this special type of entities in the brain plasma. By simply adjusting $U_{x0}$ from 0.1 to 0.3, Figure 5 reconstructs the measured γ-envelopes given in the top panel of Figure 1. Though a preliminary result it is, we see that the simulations are capable of signifying evidently the irregular occurrence of the measured stormy train of the amplitude-modulated wave packets.

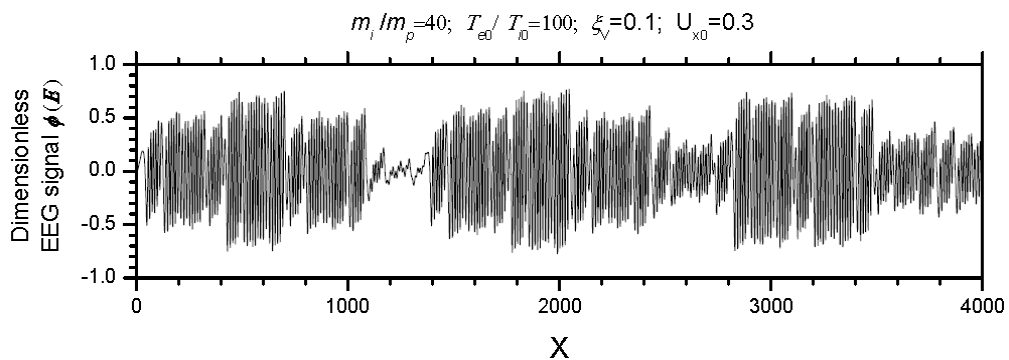

Figure 5. Simulations of complex EEG γ-signals, $\phi(E)$

---

[43] Sauer K, Dubinin E, McKenzie JF 2001. New type of soliton in bi-ion plasmas and possible implications. Geophys. Res. Lett. 28, pp.3589–3592.

5. SUMMARY AND CONCLUSION

According to the recorded wave forms, the EEG signals can be divided into two distinct groups. Group-1 owns highly nonlinear structures with a train of storm-like wave packets the amplitude of which are modulated violently to display complex envelopes. This group contains α, β, and γ types. By contrast, Group-2 is composed of quasilinear waves with deformed amplitudes deviated more or less from linear waves. This group includes θ and δ types.

Since the plasma model of brain dynamics was formulated in the early 1970s,[15] new advance has been reported toward the collision-free processes happening in the brain plasma system (e.g., ref.18). More importantly, QBD was suggested to explain the neuro- and-cognitive mechanism of human consciousness.[8] However, little work is known to elucidate the mechanism of the electric EEG signals measured in the human brain under different situations of the mental consciousness.

We derived a set of two-fluid, self-similar, nonlinear solitary wave equations from PBD's Vlasov-Maxwell equations. This model treats brain aqua-ions and electrons as two different fluids which are coupled with each other in the presence of the internal electric and magnetic fields and the external geomagnetic field. The **E**×**B** draft becomes a criterion to differentiate the effect of the plasma particles. By making use of the dimention-free formalism, we perform numerical simulations to fit with the five types of the EEG signals. Following results are obtained:

(1) In the external geomagnetic field, $B_{ext}$, the **E**×**B** speed in the brain is in the same order of the ion thermal speed, but much smaller than the electron thermal speed. Thus, brain electric field has little correlation with the dynamics of electrons.

(2) The formation of the EEG waves is dependent of not only electric and magnetic fields, but also brain aqua-ions, electron and ion temperatures, masses, and their initial fluid speeds. Different sets of these input parameters contribute to different solitary wave packets.

(3) When electron inertia is neglected, Group-2 simple waves come into being within the IA band. The waves are featured by one or a combination of the three envelopes: sinusoidal, sawtooth, and spiky/bipolar, among which the number of harmonic ingredients is different.

(4) When electron inertia is considered, Group-1 complex waves emerge as the superimpositions of following two or three components: the high-frequency spiky IA mode, the intermediate-frequency oscillating LH mode, and, the low-frequency fluctuating IC mode.

By introducing the different plasma brain model from QBD, the present paper proposes a new qualitative overview on the mechanism of the brain EEG waves. Though the waves in the brain are shown to have similar dynamical processes to those in space plasmas, quantitative studies are necessary in clinic diagnoses of measured EEG signals.

zma@mymail.ciis.edu
Philosophy, Cosmology and Consciousness
California Institute of Integral Studies
1453 Mission St., San Francisco, CA 94103